\documentclass[aps,prb,reprint,twocolumn,superscriptaddress,showkeys]{revtex4-2}
\usepackage{url}
\usepackage{amsmath}
\usepackage{amssymb}
\usepackage{graphicx}
\usepackage{dcolumn}
\usepackage{bm}
\usepackage{hyperref}

\begin{document}

\title{Parallel Simulation of Josephson Junctions With Multiplicative Noise}

\author{Vincenzo Pierro}
\email{pierro@unisannio.it}
\affiliation{Department of Engineering, University of Sannio, Benevento I-82100, Italy}

\author{Luigi Troiano}
\email{troiano@unisannio.it}
\affiliation{Department of Engineering, University of Sannio, Benevento I-82100, Italy}

\author{Elena Mejuto Villa}
\email{mejutovilla@unisannio.it}
\affiliation{Department of Engineering, University of Sannio, Benevento I-82100, Italy}

\author{Sergio Pagano}
\email{spagano@unisa.it}
\affiliation{Dipartimento di Fisica E.R. Caianiello and CNR-SPIN, University of Salerno, Fisciano I-84084, Italy}

\author{Giovanni Filatrella}
\email{filatrella@unisannio.it}
\affiliation{Department of Science and Technology, University of Sannio, Benevento I-82100, Italy}

\date{October 2018}

\begin{abstract}
Parallel graphic processing units have been employed for fast simulations of the switching dynamics of Josephson junctions subject to critical current fluctuations. Such a system is modeled by a nonequilibrium washboard model with multiplicative noise, for which analytical results are lacking. The proposed approach allows us to execute extensive numerical simulation in short time and with relatively inexpensive resources. This allows us to fully characterize the effect of the noise on the junction switching current distributions at realistic bias current sweeprates.
\end{abstract}


\maketitle

\section{Introduction}
This paper deals with effective and realistic simulations of a Josephson junction (JJ) subject to a multiplicative noise, arising if the junction critical current $I_0$, the maximum bias current that can sustain the zero voltage state, is subject to random fluctuations. In the deterministic model that neglects fluctuations, the parameter $I_0$ is the maximum current carried by the Cooper's pairs; above this current value, the junction is not anymore in the superconducting state and switches to the normal state. To determine this JJ fundamental parameter $I_0$, one usually records the current-voltage characteristics, which are obtained by sweeping the bias current and simultaneously recording the average junction voltage drop.

Usually, the presence of noise in JJs is attributed to the Johnson current noise associated to the quasiparticle tunneling resistance. Such noise is represented as an additional random bias current source in parallel with the standard dc, or low frequency, current source. The random bias leads to an additive noise term in the equations describing the dynamics of the junction [1]. Thus, when a bias current noise is included in this picture, the noise is additive and one observes that the passage to the finite-voltage state occurs at bias currents lower that $I_0$ and randomly distributed around a mean value [1]. By repeating many times the biasing sequence, a distribution of the switching currents is obtained, whose statistical properties are of great importance in many JJ applications, both in the thermal and quantum regimes [2]. The randomness affects the switching current distribution [2] of JJ detectors [3]--[7], is a disturbance in macroscopic quantum tunneling [8]--[10] and in quantum computation [11].

However, if fluctuations of the JJ critical current $I_0(t)$ are considered, the Cooper pair tunneling current is directly affected, and the resulting noise term in the model equations is multiplicative. The source of these fluctuations can be intrinsic, e.g., thermal, or external, e.g., induced by the interaction with electromagnetic fields. In such case, the nonequilibrium model of the junction contains a multiplicative noise, that cannot be analytically converted into additive noise [12], [13], and requires extensive numerical simulations. Moreover, in real experiments the bias current is usually ramped up at a rate from few hertz to few kilohertz, several orders of magnitude lower of the intrinsic junction frequency, which is of the order of few gigahertz. Thus, to perform accurate simulations of the experimental findings in the presence of noise, very long calculations are required.

Fortunately, stochastic calculus can be parallelized assigning different realizations to different process units [14]. In this paper, graphic processing units are used for such parallel calculations. This possibility has recently arisen since the availability of a platform compute unified device architecture (CUDA) [15] for general purpose computation using graphic units processors, that is both cheap and extremely fast [16]. This paper is organized as follows. In Section II, the model of JJ with multiplicative noise is derived and the numerical scheme briefly described.

In Section III, the results obtained from the numerical simulations are discussed, and finally in Secttion IV, conclusion and future perspectives are discussed.

\section{Model}
In this section, the mathematical model of a JJ under the influence of multiplicative noise is developed and the parallel algorithm implemented on CUDA is described.

\subsection{Model Equations}
The starting point is the usual model of a current biased JJ [1], where the bias current is divided into three different channels (see Fig.~1)
\begin{equation}
I_b(t) = I_R + I_C + I_J.
\end{equation}

\begin{figure}[htbp]
\centering
\includegraphics[width=\columnwidth]{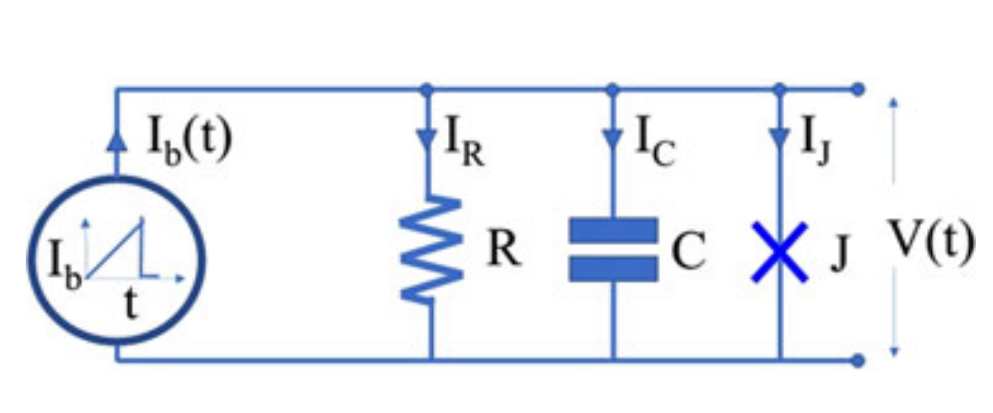}
\caption{Circuit model of a JJ with multiplicative noise. The sawtooth bias current is assumed noiseless, while the parameter $I_0$ of the JJ element randomly fluctuates as per (3).}
\label{fig:fig1}
\end{figure}

Here, $I_R$ is the normal electron current, due to the quasiparticle tunneling or to an external shunting resistor, $I_C$ is the displacement current, due to the junction capacitance, $I_J$ the Cooper pair tunneling current, and $I_b(t)$ the noiseless bias current, usually a linearly growing ramp. By using the Josephson equations [1], (1) becomes
\begin{equation}
C\frac{\hbar}{2e}\frac{d^2\varphi}{dt^2} + \frac{\hbar}{2e}\frac{1}{R}\frac{d\varphi}{dt} + I_0(t)\sin(\varphi) = I_b(t)
\end{equation}
where $e$ is the elementary charge, $\hbar$ is the reduced Planck constant, $\varphi$ the Josephson phase, and $I_0(t)$ is the Josephson critical current that, in this context, is subject to random fluctuations, and hence, explicitly depends upon the time.

Such random fluctuations may be due, for instance, to the interaction with a random electromagnetic field surrounding the junction, whose effect is to change the critical current $I_0$ around a mean value $\bar{I}_0$. To avoid unnecessary complications, it is assumed that that the critical current is subject to Gaussian fluctuations of the type
\begin{equation}
I_0(t) = \bar{I}_0 + \Delta I_0(t)
\end{equation}
where the random term $\Delta I_0$ is Gaussian correlated
\begin{equation}
\langle \Delta I_0(t) \rangle = 0
\end{equation}
\begin{equation}
\langle \Delta I_0(t)\Delta I_0(t') \rangle = 2A\delta(t - t').
\end{equation}
The above (2)--(5) can be transformed into normalized units by the scaling $\tau = \omega_J t$, with the plasma frequency $\omega_J = \sqrt{2eI_0/\hbar C}$, thereby it is obtained
\begin{equation}
\frac{d^2\varphi}{d\tau^2} + \alpha \frac{d\varphi}{d\tau} + [1 + \xi(\tau)] \sin(\varphi) = \gamma(\tau)
\end{equation}
where the normalized dissipation is $\alpha = (1/R)\sqrt{\hbar/(2eI_0C)}$, assumed to be equal to $0.05$ in all subsequent numerical simulations, and the normalized noise is characterized by
\begin{equation}
\langle \xi(\tau) \rangle = 0,
\end{equation}
\begin{equation}
\langle \xi(\tau)\xi(\tau') \rangle = 2\Sigma\delta(\tau - \tau').
\end{equation}
In (8), the noise intensity is $\Sigma = A/(\omega_J I_0^2)$. To model the experimental situation a time dependent bias is considered, that increases at a fixed speed $v_\gamma$
\begin{equation}
\gamma(\tau) = v_\gamma \tau.
\end{equation}
In the absence of noise, the switches would occur at a time $\tau = 1/v_\gamma$, corresponding to a bias current equal to the Josephson critical current ($\gamma = 1$). In the presence of noise, the switches occur at $\gamma < 1$. Naturally, the simulation time is inversely proportional to the speed $v_\gamma$. Using the following typical values of a JJ [17], [18]: critical current density $J_C = 1000\text{ A/cm}^2$, area $A = 10\ \mu\text{m}^2$, specific capacitance $C_s = 0.05\text{ F/m}^2$, and a resistance $R = 50\ \Omega$, one obtains a critical current $I_0 \sim 100\ \mu\text{A}$, a capacitance $C \sim 0.5\text{ pF}$, and a plasma frequency $\omega_J \sim 8 \times 10^{11}\text{ rad/s}$. In the inverse plasma frequency timescale, an ordinary bias current ramp with a repetition rate of $\sim 8\text{ kHz}$ would result in a normalized speed $v_\gamma \sim 10^{-8}$. Numerical simulations, therefore, require extremely long (normalized) times of the order of $\sim 10^8$ units, or $\sim 10^{11}$ time steps. This number of steps is not prohibitive, and can be performed for a single stochastic evolutions. However, simulations require to average over many realizations, and these can be conveniently implemented in parallel simulations, as will be outlined in the following section.

\subsection{Parallel Algorithm for Multiplicative Noise With CUDA on Graphical Processing Units}
To challenge realistic simulations for the measurement of JJ switching current, parallel codes running on CUDA has been employed. In particular, the algorithm developed for determining the switching currents under the effect of additive noise [16] has been adapted to the case of multiplicative noise. In this algorithm, each processor is assigned to the simulation of a single bias sweep event, in presence of noise. The switching current corresponds to an exit, or transition, time that is the time to pass a threshold, somewhat related to Kramers' escape rate [19]. This exit time is random, and therefore, the algorithms have been adapted to schedule the computational load to different processing units as soon as the processor is made available, as discussed in [16]. When the algorithm is applied to multiplicative noise, it achieves the same acceleration, with respect to single processor execution, of the order of $\sim 500$ times, as in the case of additive noise. This acceleration, that has been estimated with parallel calculations on fast processes, is the essential ingredient for simulations at realistic bias sweep rates.

\section{Numerical Results}
During each bias current ramp (called ``trial''), the presence of noise induces a premature switching of the junction to the voltage state. This is shown in Fig.~2, where the cumulative switching distribution function, computed after $100\,000$ trials with $v_\gamma = 10^{-6}$ and $\Sigma = 0.015$, is shown together with the corresponding switching current distribution. In the absence of noise, $\Sigma = 0$ in (8), the cumulative distribution function (CDF) curve in Fig.~2 would represent a $\theta(\gamma - 1)$ function, where $\theta(\cdot)$ denote the unit step function. The effect of noise is to produce a finite switching probability for $\gamma < 1$.

\begin{figure}[htbp]
\centering
\includegraphics[width=\columnwidth]{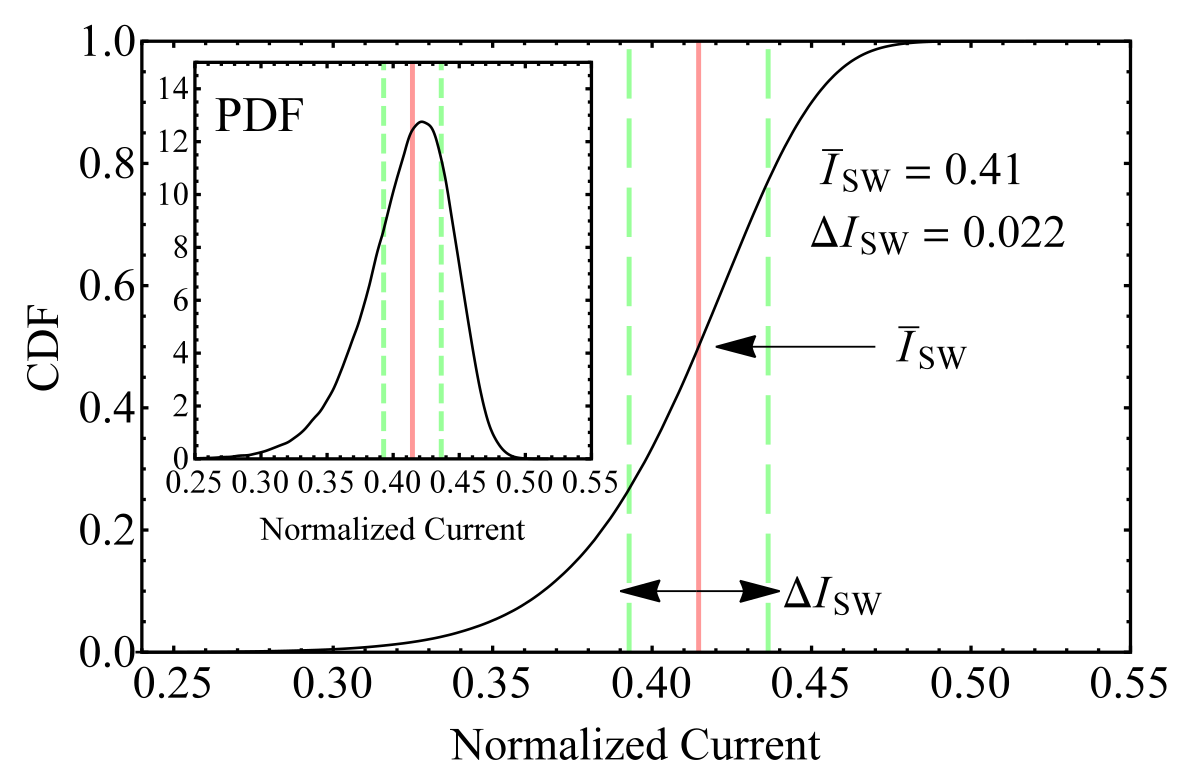}
\caption{CDF of the switching currents as a function of the bias current. The corresponding switching current distribution is shown in the inset. The values of the current distribution width $\Delta I_{\mathrm{SW}}$ and median $\bar{I}_{\mathrm{SW}}$ are indicated by vertical lines.}
\label{fig:fig2}
\end{figure}

The switching current distribution can be characterized by two parameters, the median switching current $\bar{I}_{\mathrm{SW}}$ (the value at which 50\% of the trials resulted in a finite voltage) and the width of the distribution, represented by the interquartile range $\Delta I_{\mathrm{SW}}$, the interval between the 25th and the 75th percentiles (see vertical grid lines in Fig.~2).

\begin{table}[htbp]
\caption{\label{tab:tab1}\textsc{Timing of Numerical Simulations}}
\centering
% Aumenta l'altezza e il respiro delle righe (default 1.0)
\renewcommand{\arraystretch}{1.1} 
% \begin{tabular*}{\linewidth} forza la tabella a occupare l'intera larghezza della colonna
\begin{tabular*}{0.9\linewidth}{@{\extracolsep{\fill}}|c|c|c|c|}
\hline
run \# & $\Sigma$ & $v_\gamma$ & GPU time \\
\hline\hline
1 & 0.002 & $10^{-8}$ & $6 \times 10^5\text{ sec}$ \\
\hline
2 & 0.010 & $10^{-8}$ & $3 \times 10^5\text{ sec}$ \\
\hline
3 & 0.015 & $10^{-8}$ & $1 \times 10^5\text{ sec}$ \\
\hline
4 & 0.015 & $10^{-7}$ & $1 \times 10^4\text{ sec}$ \\
\hline
\end{tabular*}
\end{table}

By using the above described approach, (6)--(8) have been numerically simulated. In the simulations the bias sweep velocity has been varied from $v_\gamma = 10^{-3}$ to $v_\gamma = 10^{-8}$, to reproduce realistic bias sweep rates. To integrate the stochastic differential equations it has been employed the Euler--Maruyama method [20] with a normalized time step of $10^{-3}$. In each trial the normalized bias current is linearly increased, starting from 0, at a rate $v_\gamma$ until a switching occurs, defined by using the Josephson phase crossing of the top of washboard potential barrier. A complete run comprises $10^5$ trials. Table~I shows some examples of the GPU time required by the numerical simulations for different noise levels and ramp speeds. Naturally, the slower is the ramp speed the longer will be the simulation run duration. The noise amplitude has a less marked effect on the simulation timing, with a general tendency to decrease with the rising of noise level. The acceleration respect to the CPU times for the cases of Table~I is estimated to be about 500 on much shorter runs [16], as direct comparison would require simulation times on sequential CPU in the order of $10^7\text{ s}$, that is several months. Overall, the results in the Table~I demonstrate that GPU calculations can bring the computing time for stochastic simulations [16] within feasible limits. The simulations have been performed on a desktop computer equipped with a NVIDIA Titan Xp GPU. The NVIDIA Titan Xp card is based on the GP102 GPU, clocked at 1481 MHz and able to reach 12.1 TFLOPs. The on-board memory is 12 GB GDDR5X with a 384-bit bus width and 11.4 Gb/s bandwidth.

\begin{figure}[htbp]
\centering
\includegraphics[width=\columnwidth]{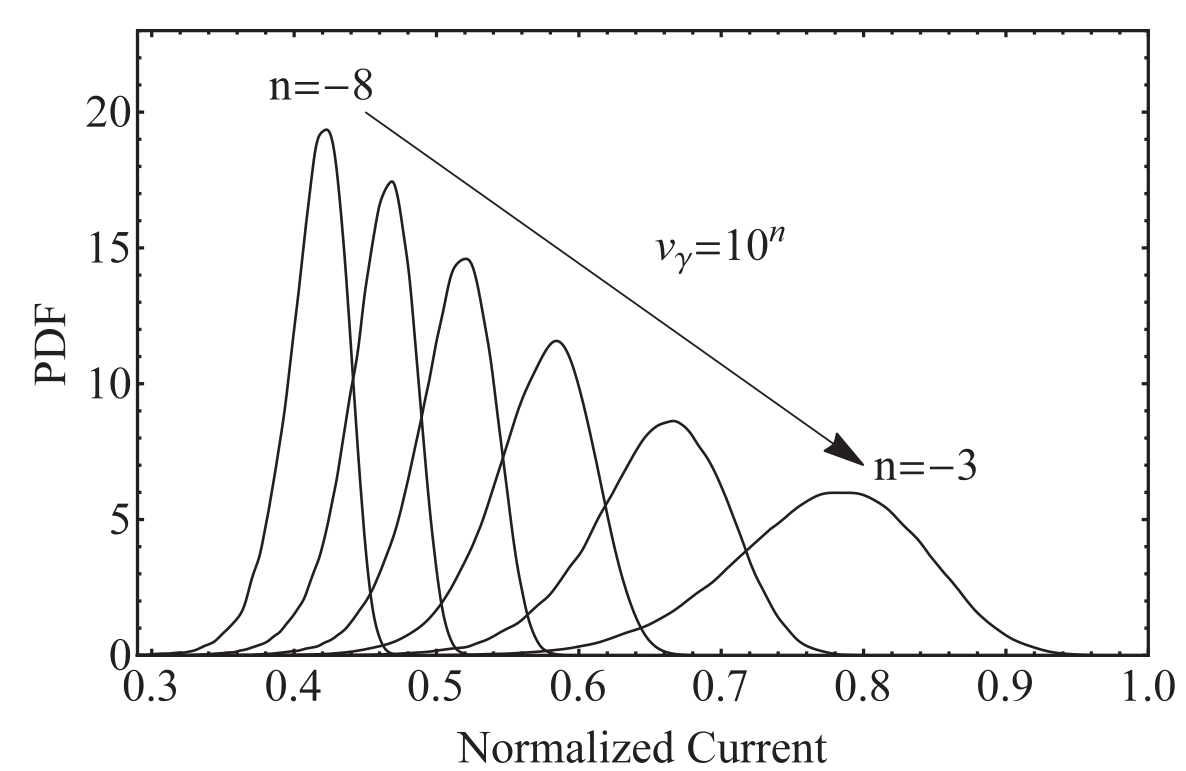}
\caption{Simulation results of the distributions of the current switches for increasing values of the current ramp speed $v_\gamma$ for $\Sigma = 0.010$.}
\label{fig:fig3}
\end{figure}

In each simulation run, the recorded switching current values are used to compute the probability distribution function (PDF), using a kernel density estimation [5], [6]. The effect of the bias sweep velocity on the shape of the PDF is shown in Fig.~3. The curves in the figure show that, in the case of multiplicative noise, the effect of the bias sweep velocity is to shift the distributions toward higher current values, while at the same time increasing their width. A similar, albeit less pronounced, can be seen in the ordinary additive thermal noise models [2]. This can be understood by noting that, at a slower ramp speeds, large current fluctuations can occur more frequently, and induce a switching at lower bias currents. Additionally, at fast ramp speeds, the dynamic effect of changing the washboard potential during the phase evolution is also present.

\begin{figure}[htbp]
\centering
\includegraphics[width=\columnwidth]{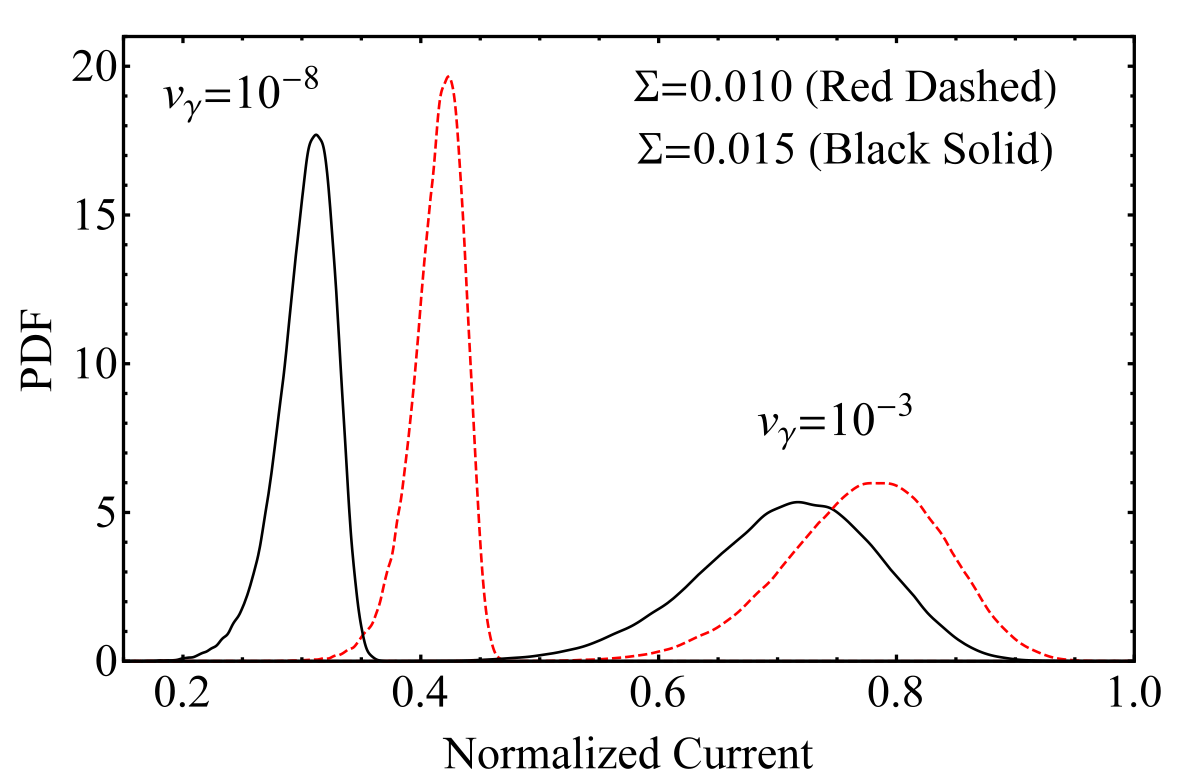}
\caption{Simulation results as a function of the multiplicative noise intensity $\Sigma$ for two values of the current ramp speed, $v_\gamma = 10^{-8}$ and $v_\gamma = 10^{-3}$.}
\label{fig:fig4}
\end{figure}

Another relevant result of numerical simulations is the effect of the multiplicative noise intensity $\Sigma$ on the shape of the distributions, as shown in Fig.~4. Also, in this case, the results indicate that multiplicative noise has an effect similar to additive noise: A decrease of the noise intensity shifts the distribution toward higher current.

Although the type of noise considered in this paper is not related to thermal noise, it can be interesting to introduce an effective temperature, by fitting the shape of the computed distribution with that predicted by the adiabatic additive noise approach [2]. In Table~II, the values of the equivalent temperature $T_{\mathrm{fit}}$ for different values of the noise amplitude and bias ramp speed are reported. From the table it is evident that the equivalent temperature increases with the increase of the multiplicative noise intensity. However, in contrast with what is predicted by thermal models, for the same noise amplitude the equivalent temperature depends on the ramp speed, as in a quantum model [21]. Such result is due to the intrinsically different effect of the multiplicative noise on the escape dynamics and underlines the necessity to better understand the way the various noise sources act on Josephson devices.

\begin{table}[htbp] % Usa [H] per bloccarla esattamente qui
\caption{\label{tab:tab2}\textsc{Comparison of Additive and Multiplicative Noise}}
\centering
\renewcommand{\arraystretch}{1.35} 
\begin{tabular*}{0.9\linewidth}{@{\extracolsep{\fill}}|c|c|c|c|}
\hline
$\Sigma$ & $v_\gamma$ & $\gamma_{\text{peak}}$ & $T_{\text{fit}}$ \\
\hline\hline
0.015 & $10^{-8}$ & 0.31 & 1.30 \\
\hline
0.015 & $10^{-3}$ & 0.71 & 0.47 \\
\hline
0.010 & $10^{-8}$ & 0.42 & 0.97 \\
\hline
0.010 & $10^{-3}$ & 0.78 & 0.30 \\
\hline
\end{tabular*}
\end{table}

\section{Conclusion}
We have numerically investigated the switching current distributions in JJs under the effect of multiplicative noise. Such type of noise can arise from fluctuations of the junction critical current, due to intrinsic (e.g., thermal or quantum) or external (electromagnetic fields) causes. This stochastic dynamic model has been solved by parallel numerical simulations on graphical processing units. The speedup obtained by parallel processing has allowed to extend the simulations to realistic current ramp speeds. The results show differences and similarities with the extensively studied case of additive noise. An attempt to introduce the thermal equivalent of multiplicative noise is also reported.

The speed, and the low cost of processing on graphical units allows us to extend this paper to large systems consisting of many JJs, as might be required for high-Tc superconductors [22] or extended JJ [23].

The nonlinear stochastic dynamical system investigated is quite generic and can be applied to model a vast class of systems, apart from JJs, e.g., Atomic Force Spectroscopy diagnostics [24], nanowires [25], power grids [26], and challenging detection problems [27], [28]. Moreover, as there is some controversy about the Kramers' rate of a multiplicative noise [13], an appropriated numerical method for extensive simulations, alongside with experiments, can be useful.

\begin{acknowledgments}
The authors would like to thank the support of NVIDIA Corporation with the donation of the Titan Xp GPU used for this research.
\end{acknowledgments}

\end{document}